\documentclass[showpacs,twocolumn,groupedaddress,superscriptaddress,aps,prb]{revtex4}

\usepackage[greek,english]{babel}
\usepackage{graphicx}
\usepackage{amsmath}
\usepackage{mathrsfs}
\usepackage{amsfonts}
\usepackage{amssymb}
\usepackage[utf8]{inputenc}
\usepackage{booktabs}
\usepackage{float}

\usepackage[colorlinks=true,bookmarks=false,citecolor=blue,urlcolor=blue,pdfstartview=FitH]{hyperref}

\begin{document}

\title{Effect of loss on the dispersion relation of photonic and phononic crystals}

\author{Vincent Laude}
\email{vincent.laude@femto-st.fr}
\affiliation{Institut FEMTO-ST, Universit\'e de Franche-Comt{\'e} and CNRS, Besan\c{c}on, France}
\author{Jose Maria Escalante}
\affiliation{Nanophotonics Technology Center, Universitat Politècnica de Valencia, Valencia 46022, Spain}
\author{Alejandro Mart{\'i}nez}
\affiliation{Nanophotonics Technology Center, Universitat Politècnica de Valencia, Valencia 46022, Spain}

\begin{abstract}
A theoretical analysis is made of the transformation of the dispersion relation of waves in artificial crystals under the influence of loss, including the case of photonic and phononic crystals.
Considering a general dispersion relation in implicit form, an analytic procedure is derived to obtain the transformed dispersion relation.
It is shown that the dispersion relation is generally shifted in the complex $(k,\omega)$ plane, with $k$ the wavenumber and $\omega$ the angular frequency.
The value of the shift is obtained explicitly as a function of the perturbation of material constants accounting for loss.
The method is shown to predict correctly the transformation of the complex band structure $k(\omega)$.
Several models of the dispersion relation near a symmetry point of the Brillouin zone are analyzed.
A lower bound for the group velocity, related to the local shape of the band around symmetry points, is derived for each case.
\end{abstract}

\maketitle

\section{Introduction}

Wave propagation in heterogeneous artificial crystals composed of two or more materials that differ in their material constants has received great attention during the last two decades.
Artificial crystals include in particular the cases of photonic crystals for optical waves \cite{joannopoulosBOOK2008} and of phononic crystals for elastic waves~\cite{kushwahaPRL1993,pennecSSR2010}.
Thanks to advances in nano-fabrication, photonic crystals in the visible and near IR, and phononic crystals in the gigahertz spectral range are now available.
They can be used to manipulate, to guide, and to confine light or sound, or even both type of waves at same time (phoxonic crystals) \cite{maldovanAPL2006,sadatJAP2009}.
One important property of this type of structure is slow wave propagation \cite{babaNP2008,thevenazNP2008,laudeOE2011}.
Many devices may potentially take advantage of slow wave effects, e.g., for amplification, optical buffering, acousto-optic modulation, sensing, pulse compression, or enhanced non-linear effects. 
Numerous experimental results, mainly in the field of photonic crystals, have however demonstrated that despite tremendous fabrication efforts, reducing the group velocity beyond two orders of magnitude still remains elusive.\cite{vlasovN2005}
One limitation imposed on the minimum attainable group velocity of both photonic and phononic crystals has its origin in propagation losses.

Losses can be due to fabrication disorder or to intrinsic material losses.\cite{hughesPRL2005,ofaolainOE2007,pedersenPRB2008,husseinPRB2009,moiseyenkoPRB2011}
There exist many different models of loss, that are adapted to different situations.
With Helmholtz-type propagation equations, obtained with the assumption of monochromatic waves, it is convenient to introduce loss via the imaginary part of material constants that are implicitly dependent on frequency.
In the photonic crystal case, the position dependent dielectric tensor is written as\cite{pedersenPRB2008}
\begin{equation}
\epsilon(\mathbf{r}) = \epsilon'(\mathbf{r}) - \imath \epsilon''(\mathbf{r}) .
\label{eq1}
\end{equation}
In the phononic case, the elastic tensor is written as \cite{moiseyenkoPRB2011}
\begin{equation}
C(\mathbf{r}) = C'(\mathbf{r}) + \imath C''(\mathbf{r}) .
\label{eq2}
\end{equation}
In either case, the real part of the material constants is the value in the absence of loss and the imaginary part is assumed to be relatively small, i.e., $\epsilon''/\epsilon' << 1$ and $C''/C' << 1$.
In addition, the imaginary part can depend on frequency.
This dependence is explicit in viscoelastic models where $C'' = \eta \omega$ with $\eta$ the viscoelasticity tensor and $\omega$ the angular frequency.
The sign of the imaginary parts in (\ref{eq1}) and (\ref{eq2}) is consistent with the time-harmonic dependence $\exp(\imath(\omega t - \mathbf{k} \cdot \mathbf{r}))$ where $\mathbf{k}$ is the wavevector.

The transformation of dispersion relations when loss is included is a general question with both theoretical and practical implications.
With homogeneous media, the analysis is straightforward.
In artificial crystals, however, periodicity introduces a strong dispersion that makes the problem more intricate.
This is especially true at degeneracy points of the dispersion relation where the group velocity becomes very small or even vanishes.
Pedersen \textit{et al.} studied how material losses induce a lower bound for the group velocity in photonic crystals.\cite{pedersenPRB2008}
The method of analysis considers an approximate model of the dispersion relation near a band edge, at a symmetry point of the first Brillouin zone.
The exact shape of the dispersion relation can quite complex, but in the vicinity of a point where the group velocity vanishes in the absence of loss, the band structure can be approximated as
\begin{equation}
\omega = \omega_0 + \alpha (k-k_0)^2 .
\label{eq3}
\end{equation}
The wavenumber $k$ is here a component of the wavevector in a particular propagation direction.
From first-order perturbation theory, Pedersen \textit{et al.} obtained that when loss is added
\begin{equation}
v_g \geq \sqrt{|\alpha| \omega F \frac{\epsilon ''}{\epsilon '}}
\label{eq4}
\end{equation}
where $F$ is the filling fraction.
This result is quite general and is in particular independent of the details of the photonic crystal lattice and of its composition.
For the case of phononic crystals, no similar result has been found so far, to the best of our knowledge.
Moiseyenko and Laude investigated numerically the transformation of the complex band structure when loss is introduced and obtained a expression for the group velocity at any point of the dispersion relation.\cite{moiseyenkoPRB2011}
Although the result is general and independent of the details of the phononic crystal, the limitation of the group velocity by loss was a numerical observation but was not given an explicit expression.

In this paper, we attempt to give a general theory of the influence of loss on dispersion relations in artificial crystals, encompassing and extending the results above.
Considering implicit rather than explicit dispersion relations proves to be an adequate way of tackling the problem.
In section \ref{sec2}, we show how the classical and the complex band structures can be cast in implicit form, and how the first variations of frequency and wavenumber can be estimated when loss is introduced as a perturbation of material constants.
In section \ref{sec3}, we apply the theory to different approximate models of dispersion relations, first following the typology introduced by Figotin and Vitebskiy \cite{figotinWRCM2006} and then introducing a model of the complex dispersion relation in a Bragg band gap.
In each case, an analytical expression of the lower bound set by loss on the group velocity is given.
Our results hold in general for a Helmholtz-type equation (a monochromatic wave equation), including periodicity and anisotropy.

\section{Perturbation of dispersion relations}
\label{sec2}

\subsection{Dispersion relations}

A general dispersion relation has the implicit form
\begin{equation}
{\cal D}(\omega, k, \mu) = 0
\label{eq5}
\end{equation}
where $\mu$ represents any material constant.
A band structure is a plot of a set of such dispersion relations for fixed $\mu$.
The first differential of the dispersion relation is
\begin{equation}
\frac{\partial {\cal D}}{\partial \omega} \mathrm{d}\omega + \frac{\partial {\cal D}}{\partial k} \mathrm{d}k + \frac{\partial {\cal D}}{\partial \mu} \mathrm{d}\mu = 0 .
\label{eq6}
\end{equation}
The group velocity (at fixed $\mu$, i.e., for a given set of material constants) can then be obtained as
\begin{equation}
v_g = \left. \frac{\mathrm{d}\omega}{\mathrm{d}k} \right|_{\mu} = - \frac{\frac{\partial {\cal D}}{\partial k}}{\frac{\partial {\cal D}}{\partial \omega}} .
\label{eq7}
\end{equation}
Similarly, we can look at the variation of angular frequency (at fixed $k$) caused by a variation in material constants as
\begin{equation}
\left. \frac{\mathrm{d}\omega}{\mathrm{d}\mu} \right|_{k} = - \frac{\frac{\partial {\cal D}}{\partial \mu}}{\frac{\partial {\cal D}}{\partial \omega}}
\label{eq8}
\end{equation}
and the converse variation of $k$ (at fixed $\omega$) caused by a variation in material constants as
\begin{equation}
\left. \frac{\mathrm{d}k}{\mathrm{d}\mu} \right|_{\omega} = - \frac{\frac{\partial {\cal D}}{\partial \mu}}{\frac{\partial {\cal D}}{\partial k}} .
\label{eq9}
\end{equation}
All these relations are valid for infinitesimal variations of the variables of the dispersion relation and hold exactly on it, i.e., at $(\omega, k, \mu)$ points where (\ref{eq5}) is satisfied.

\begin{figure}[!t]
\includegraphics[width=80mm]{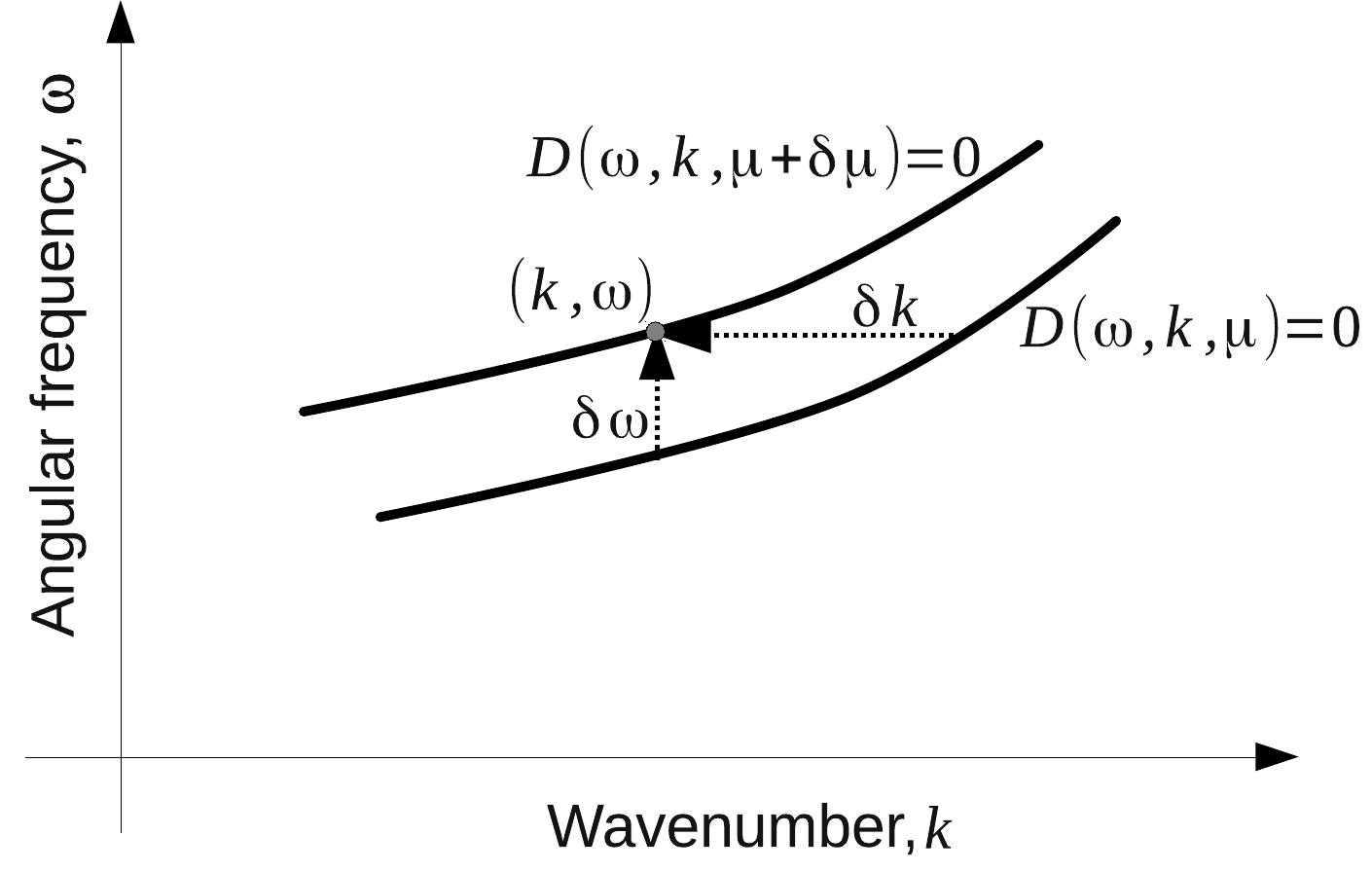}
\caption{Schematic representation of a dispersion relation in the wavenumber-frequency plane and its transformation under a perturbation of the material constants $\mu$. A point $(k, \omega)$ on the perturbed (lossy) dispersion relation can be traced backward to the initial (lossless) dispersion relation by either a vertical shift $-\delta \omega$ or a horizontal shift $- \delta k$.}
\label{fig1}
\end{figure}

Suppose now we solve for the dispersion relation for a given set of material constants $\mu$ and then for a slightly perturbed set $\mu + \delta\mu$, e.g., without/with including loss.
In practice, we would first solve ${\cal D}(\omega, k, \mu) = 0$, then ${\cal D}(\omega, k, \mu+\delta\mu) = 0$.
As figure \ref{fig1} illustrates, the distance between the initial dispersion relation and the perturbed one is $\delta \omega$ in the vertical direction and $\delta k$ in the horizontal direction.
For a small perturbation $\delta\mu$, the two dispersion relations are almost parallel in the $(k, \omega)$ plane, as expressed by the two equalities
\begin{eqnarray}
{\cal D}(\omega, k, \mu + \delta \mu) &=& {\cal D}(\omega - \delta \omega|_k, k, \mu) 
\label{eq10} \\
&=& {\cal D}(\omega, k - \delta k|_\omega, \mu) ,
\label{eq11}
\end{eqnarray}
as can be readily obtained by considering first-order Taylor expansions around the point $(\omega, k, \mu)$ in each variable and using (\ref{eq6}).
They mean that we can go from one point $(k, \omega)$ of the perturbed dispersion relation to the initial dispersion relation by a backward shift along either axis of the band structure.
They are the basic ingredients that are needed to complexify dispersion relations when loss is added, as we illustrate in Section \ref{sec3}.
Each equality is then valid only if the first derivatives with respect to the involved variable does not vanish.

\subsection{Band structure, $\omega(k)$}

The master equations for electromagnetic waves in dielectric media, and hence in photonic crystals, are of the form
\begin{equation}
\nabla \wedge \left(\frac{1}{\epsilon} \nabla \wedge \mathbf{H} \right) = \left( \frac{\omega}{c} \right)^2 \mathbf{H}
\label{eq12}
\end{equation}
where $\mathbf{H}$ is the magnetic field vector and $c$ is the celerity of light in a vacuum.
For elastic waves in solids, and hence in phononic crystals, the elastodynamic equations are
\begin{equation}
- \nabla \cdot \left( C : \nabla \mathbf{u} \right) = \omega^2 \rho \mathbf{u}
\label{eq13}
\end{equation}
where $\mathbf{u}$ is the displacements vector and $\rho$ is the mass density.

In order to obtain a numerical solution, the unknown fields are expanded over a given functional basis, e.g., Fourier exponentials for the plane wave expansion (PWE) method, or piecewise continuous polynomials for the finite element method (FEM).
As a result of a Galerkin type procedure, we end up with an eigenvalue problem of the form
\begin{equation}
A(k) \mathbf{x} = \omega^2 B \mathbf{x}
\label{eq14}
\end{equation}
where $\mathbf{x}$ is the vector of the unknown coefficients in the expansion, and $A$ and $B$ are square and symmetric matrices that are formed from material constants and expansion functions.
Periodicity is accounted for via the use of the Bloch-Floquet theorem.
Specifically, writing the expansion as $\mathbf{H}(\mathbf{r},t)$ or $\mathbf{u}(\mathbf{r},t) = \sum_n x_n \mathbf{f}_n(\mathbf{r}) \exp(\imath (\omega t - \mathbf{k} \cdot \mathbf{r}))$ with ($\mathbf{f}_n(\mathbf{r})$) a basis of periodic functions and denoting $\bar{\mathbf{f}}_n = \mathbf{f}_n \exp(- \imath \mathbf{k} \cdot \mathbf{r}))$, the form of the matrix elements is given in Table~\ref{tab1}.

\begin{table}[!t]
\caption{Expression of matrix elements for the $\omega(k)$-type dispersion relations for photonic and phononic crystals.
The notation $<|\cdots|>$ indicates integration over the unit-cell of the crystal.}
\label{tab1}
\centering
\begin{tabular}{lcc}
Crystal type & $A_{mn}(k)$ & $B_{mn}$ \\ \hline
Phononic & $<\nabla \bar{\mathbf{f}}_m | C | \nabla \bar{\mathbf{f}}_n>$ & $< \bar{\mathbf{f}}_m | \rho | \bar{\mathbf{f}}_n>$ \\
Photonic & $<\nabla \wedge \bar{\mathbf{f}}_m | \frac{1}{\epsilon} | \nabla \wedge \bar{\mathbf{f}}_n>$ & $< \bar{\mathbf{f}}_m | \frac{1}{c^2} | \bar{\mathbf{f}}_n>$ \\
\end{tabular}
\end{table}

From the eigenvalue problem, a scalar dispersion relation for each eigenvalue $\omega$ can be obtained by left-multiplying with the left-eigenvector $\mathbf{y}$.
The left-eigenvector is a right-eigenvector of the transposes of matrices $A$ and $B$.
Generally, $\mathbf{x}$ and $\mathbf{y}$ associated to the same eigenvalue are different, unless the matrices are symmetric \cite{thurstonSU1977}, which is the case here.
We thus obtain the dispersion relation in implicit form as
\begin{equation}
{\cal D}(\omega, k, \mu) = \mathbf{x}^T A(k) \mathbf{x} - \omega^2 \mathbf{x}^T B \mathbf{x}
\label{eq15}
\end{equation}
where the subscript $^T$ denotes transposition.

Note that the eigenvector depends on $\omega, k$, and $\mu$.
Because it satisfies (\ref{eq14}) along the dispersion relation, however, we obtain
\begin{eqnarray}
\frac{\partial {\cal D}}{\partial \omega} &=& \mathbf{x}^T \frac{\partial A(k)}{\partial \omega} \mathbf{x}  - 2 \omega \mathbf{x}^T B \mathbf{x} , \\
\frac{\partial {\cal D}}{\partial k} &=&  \mathbf{x}^T \frac{\partial A(k)}{\partial k} \mathbf{x} , \\
\frac{\partial {\cal D}}{\partial \mu} &=&  \mathbf{x}^T \frac{\partial A(k)}{\partial \mu} \mathbf{x} - \omega^2 \mathbf{x}^T \frac{\partial B}{\partial \mu} \mathbf{x} ,
\end{eqnarray}
at every point of the dispersion relation.
Since we know explicit expressions for all matrices, these formulas allow for a very efficient evaluation of the group velocity and of the variations by (\ref{eq7}--\ref{eq9}) at every $(k, \omega)$ point of the dispersion relation, given only the knowledge of the eigenvalue and eigenvector.
For instance, the group velocity can be calculated without any approximation along the dispersion relation by the formula
\begin{equation}
v_g = \frac{\mathbf{x}^T \frac{\partial A(k)}{\partial k} \mathbf{x}}{2 \omega \mathbf{x}^T B \mathbf{x} - \mathbf{x}^T \frac{\partial A(k)}{\partial \omega} \mathbf{x}} .
\label{eq18}
\end{equation}

Similarly, when the material constants are changed by a very small amount, the first variation of $\omega$ at constant $k$ and the first variation of $k$ at constant $\omega$ can respectively be estimated as
\begin{eqnarray}
\delta \omega |_k &=& \frac{\mathbf{x}^T \delta A(k) \mathbf{x} - \omega^2 \mathbf{x}^T \delta B \mathbf{x}}{2 \omega \mathbf{x}^T B \mathbf{x} - \mathbf{x}^T \frac{\partial A(k)}{\partial \omega} \mathbf{x}} , \\
\delta k |_\omega &=& - \frac{\mathbf{x}^T \delta A(k) \mathbf{x} - \omega^2 \mathbf{x}^T \delta B \mathbf{x}}{\mathbf{x}^T \frac{\partial A(k)}{\partial k} \mathbf{x}} .
\label{eq20}
\end{eqnarray}
A particularly appealing formula can be obtained for the first variation $\delta \omega |_k$ for the case of loss.
For both photonic and phononic problems, $\delta B=0$. Furthermore, $\mathbf{x}^T \frac{\partial A(k)}{\partial \omega} \mathbf{x}$ is generally negligible as compared to $2 \omega \mathbf{x}^T B \mathbf{x}$, because the former term originates only from a possible dependence of loss with frequency.
In the photonic case, we have
\begin{equation}
\omega^2 \mathbf{x}^T B \mathbf{x} = \mathbf{x}^T A \mathbf{x} = <\nabla \wedge \mathbf{H} | \epsilon^{-1} | \nabla \wedge \mathbf{H} >
\label{eq22}
\end{equation}
and
\begin{equation}
\mathbf{x}^T \delta A \mathbf{x} = <\nabla \wedge \mathbf{H} | \imath \epsilon'' / \epsilon'^2 | \nabla \wedge \mathbf{H} > .
\label{eq23}
\end{equation}
The bra-ket notation $<|\cdots|>$ indicates integration over the unit-cell of the crystal.
Considering the case of a photonic crystal of air holes in a dielectric material, we define the loss factor $L=\epsilon''/\epsilon'$ and the filling fraction
\begin{equation}
F = \frac{<\nabla \wedge \mathbf{H} | \epsilon'^{-1} | \nabla \wedge \mathbf{H} >_\mathrm{dielectric}}{<\nabla \wedge \mathbf{H} | \epsilon'^{-1} | \nabla \wedge \mathbf{H} >_\mathrm{crystal}} .
\label{eq24}
\end{equation}
The filling fraction is a dimensionless number smaller than 1 measuring the proportion of optical energy inside the dielectric as compared to the total optical energy.
Since the loss factor is assumed constant within the dielectric, we obtain at once
\begin{equation}
\delta \omega |_k = \imath \frac{\omega}{2} F L .
\label{eq25}
\end{equation}
The same equation is obtained for a phononic crystal of lossless inclusions in a lossy matrix, with the definitions $L= C'' / C'$ and
\begin{equation}
F = \frac{<\nabla \mathbf{u} | C' | \nabla \mathbf{u} >_\mathrm{matrix} }{ <\nabla \mathbf{u} | C' | \nabla \mathbf{u} >_\mathrm{crystal}}.
\end{equation}
In the case of air holes, $F=1$ since all elastic energy is concentrated in the matrix.
Equation (\ref{eq25}) was obtained previously by Pedersen et al. \cite{pedersenPRB2008} for photonic crystals and is here seen to hold for a general Helmholtz equation.
It has a very simple structure and periodicity is not immediately apparent in it.
As Table \ref{tab1} shows, the wavevector enters the filling fraction value through the spatial derivatives of the expansion functions $\bar{\mathbf{f}}_n = \mathbf{f}_n \exp(- \imath \mathbf{k} \cdot \mathbf{r}))$.
Because of the spatial averages over the unit-cell of the crystal, $F$ does not vary much with frequency.
Hence $\delta \omega |_k / \omega$ is almost a constant if the loss factor is independent of frequency\footnote{Within the frame of the viscoelastic model, $\delta \omega |_k / \omega = \imath F L / 2$ is almost a linear function of frequency.}.

A similar derivation for $\delta k |_\omega$ could be attempted from (\ref{eq20}).
It would not lead, however, to a result as simple as (\ref{eq25}), since matrix $A(k)$ is a second degree polynomial in $k$.
Furthermore, at dispersion points where $v_g=0$, $\mathbf{x}^T \frac{\partial A(k)}{\partial k} \mathbf{x}=0$ from (\ref{eq18}), and hence $\delta k |_\omega$ is seen to diverge from (\ref{eq20}).
Thus, at such points the first-order perturbation result collapses for $\delta k |_\omega$.
In contrast, $\delta \omega |_k / \omega$ is always small, which means that the first-order perturbation result is valid whatever the frequency.
In Section \ref{sec3}, we will discuss how knowledge of the local topology of the dispersion relation can be used to infer $\delta k |_\omega$ from $\delta \omega |_k$.

\subsection{Complex band structure, $k(\omega)$}

Complex band structures can be found by searching for the wavenumber as a function of frequency, as demonstrated for photonic~\cite{hsuePRB2005} and phononic~\cite{laudePRB2009,laudeAIP2011} crystals.
Whether the approach relies on the extended plane wave expansion (EPWE) or on FEM, a generalized eigenvalue problem of the form
\begin{equation}
C(\omega) \mathbf{x} = k D(\omega) \mathbf{x}
\end{equation}
is obtained.
Matrices $C$ and $D$ account for periodicity where required, are non symmetric in general, and implicitly depend on the material constants.
For brevity, we don't give the detailed expression of the matrices, but they can be found in the quoted references.

As before, we can obtain an implicit dispersion relation for each particular eigenvalue $k$, by left-multiply by the left-eigenvector $\mathbf{y}$.
The left-eigenvector satisfies
\begin{equation}
\mathbf{y}^T C(\omega) = k \mathbf{y}^T D(\omega) .
\end{equation}
The dispersion relation in implicit form is thus
\begin{equation}
{\cal D}(\omega, k, \mu) = \mathbf{y}^T C(\omega) \mathbf{x} - k \mathbf{y}^T D(\omega) \mathbf{x}
\end{equation}
whose first differentials when evaluated on the dispersion relation are
\begin{eqnarray}
\frac{\partial {\cal D}}{\partial \omega} &=& \mathbf{y}^T \frac{\partial C}{\partial \omega} u - k \mathbf{y}^T \frac{\partial D}{\partial \omega} \mathbf{x} , \\
\frac{\partial {\cal D}}{\partial k} &=& - \mathbf{y}^T D \mathbf{x} , \\
\frac{\partial {\cal D}}{\partial \mu} &=& \mathbf{y}^T \frac{\partial C}{\partial \mu} \mathbf{x} - k \mathbf{y}^T \frac{\partial D}{\partial \mu} \mathbf{x} .
\end{eqnarray}
Again, since we know the explicit expressions for all matrices, these formulas allow for a very efficient evaluation of the group velocity and the variations by (\ref{eq7}--\ref{eq9}) at every $(\omega, k)$ point of the dispersion relation, given only the knowledge of the eigenvectors $\mathbf{y}$ and $\mathbf{x}$.
For instance, we recover the group velocity expression of Ref.~\cite{moiseyenkoPRB2011}
\begin{equation}
v_g = \frac{\mathbf{y}^T D \mathbf{x}}{\mathbf{y}^T \frac{\partial C}{\partial \omega} \mathbf{x} - k \mathbf{y}^T \frac{\partial D}{\partial \omega} \mathbf{x}} .
\end{equation}
Similarly, when the material constants are changed by a very small amount, the first variation of $\omega$ at constant $k$ and the first variation of $k$ at constant $\omega$ can respectively be estimated as
\begin{eqnarray}
\delta \omega |_k &=& - \frac{\mathbf{y}^T \delta C \mathbf{x} - k \mathbf{y}^T \delta D \mathbf{x}}{\mathbf{y}^T \frac{\partial C}{\partial \omega} \mathbf{x} - k \mathbf{y}^T \frac{\partial D}{\partial \omega} \mathbf{x}} , \\
\delta k |_\omega &=& \frac{\mathbf{y}^T \delta C \mathbf{x} - k \mathbf{y}^T \delta D \mathbf{x}}{\mathbf{y}^T D \mathbf{x}} .
\end{eqnarray}
As previously, we could explicit further these variations, but the result would not be different from the previous subsection since the classical band structure and the complex band structure are just but two representations of the same physical reality.

\section{Special dispersion relation models}
\label{sec3}

\subsection{Local models}

In this section, we make use of the tools developed in Section \ref{sec2} to obtain analytical results given only the knowledge of the shape of the dispersion relation around a given point.
Around any point $(k_0,\omega_0)$ of the dispersion relation where the group velocity doesn't vanish, a local approximation limited to first order can be considered as
\begin{equation}
{\cal D}(\omega, k, \mu) = \omega - \omega_0 - \alpha (k - k_0) = 0 .
\end{equation}
Parameter $\alpha$ here equals the group velocity $v_g$, as can be checked by direct application of (\ref{eq7}).
Under an arbitrary but small perturbation of the material constants, the transformed dispersion relation is given by application of (\ref{eq10}) or (\ref{eq11}).
Both equalities apply here since $\frac{\partial \cal{D}}{\partial \omega} \neq 0$ and $\frac{\partial \cal{D}}{\partial k} \neq 0$.
Specifically,
\begin{equation}
-\delta \omega = \alpha \delta k .
\end{equation}
Thus, from (\ref{eq25}), we obtain at once that
\begin{equation}
\delta k = - \imath \frac{1}{\alpha} \frac{\omega}{2} F L .
\end{equation}
The spatial decay thus varies linearly with loss and is inversely proportional to the group velocity, i.e., flat bands in the band structure are more sensitive to loss as compared to steep bands, as expected.

\begin{figure}[!t]
\centering
\includegraphics[width=85mm]{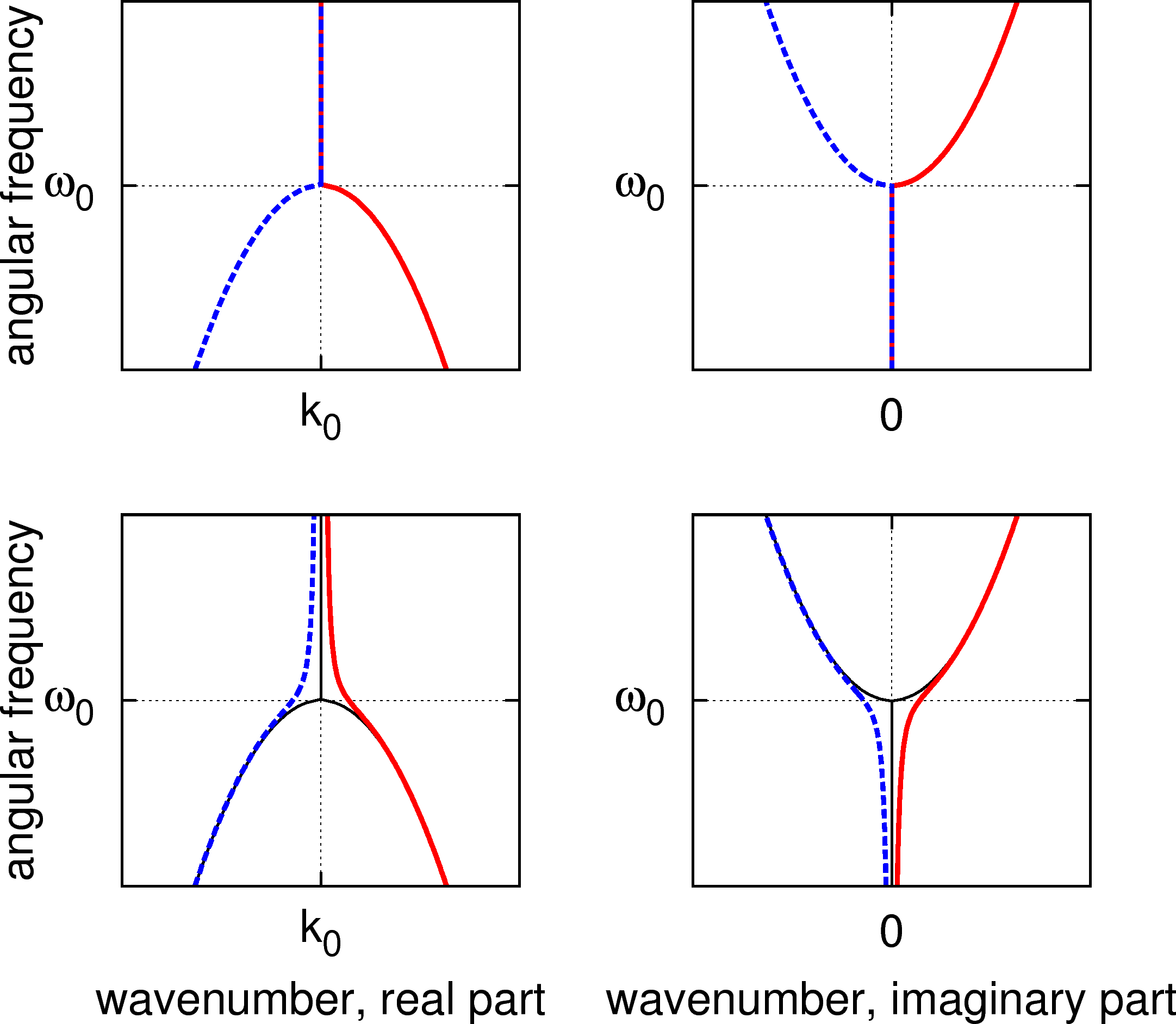}
\caption{Complex band structure $k(\omega)$ in the vicinity of a regular band edge (RBE) in the dispersion relation of an artificial crystal. Parameters for the plot are $n=2$, $\omega_0=1$, $k_0=0.5$, and $\alpha=-1$. (a) Without loss, $k(\omega)$ is real-valued below the cut-off frequency $\omega_0$ and imaginary-valued above it. The 2 bands are symmetrical with respect to $k=k_0$. (b) With loss, here with $\delta \omega / \omega_0=0.01 \imath$, $k(\omega)$ becomes complex but the 2 bands are still symmetrical with respect to $k_0$. The right-propagating (left-propagating, respectively) Bloch waves is shown with a blue line (red line, resp.).}
\label{fig6}
\end{figure}

\begin{figure}[!t]
\centering
\includegraphics[width=85mm]{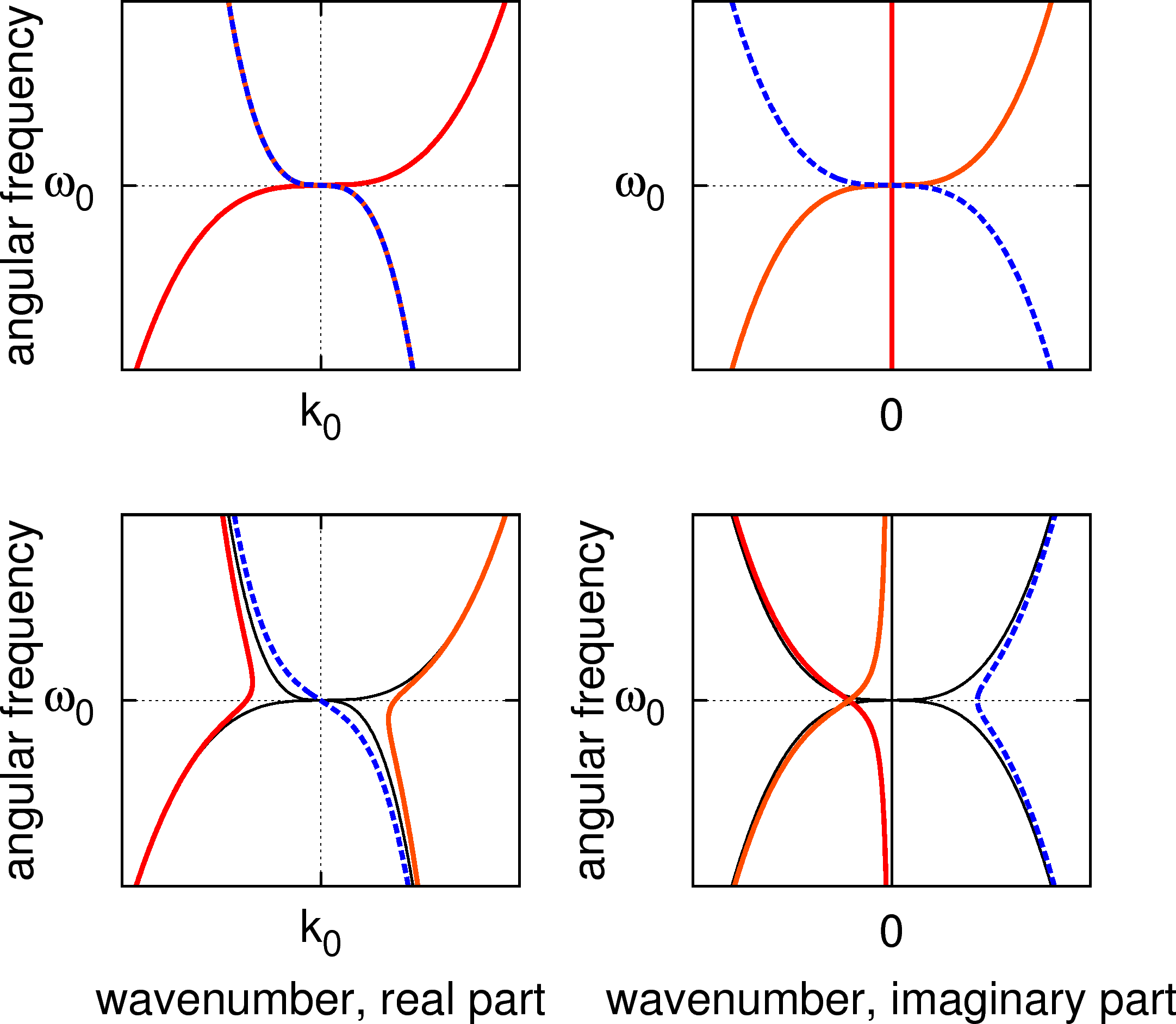}
\caption{Complex band structure $k(\omega)$ in the vicinity of a stationary inflection point (SIP) in the dispersion relation of an artificial crystal. Parameters for the plot are $\omega_0=1$, $k_0=0.5$, and $\alpha=1$. (a) Without loss, $k(\omega)$ separates in 3 bands, one is purely real, the two others are complex conjugates. (b) With loss, here with $\delta \omega / \omega_0=0.01 \imath$, $k(\omega)$ has 3 complex bands. One is a left-propagating Bloch wave (blue line), the two others are right-propagating Bloch waves (red and orange lines). Note the non continuity of band sorting at the SIP when passing from the lossless case to the lossy case.}
\label{fig7}
\end{figure}

\begin{figure}[!t]
\centering
\includegraphics[width=85mm]{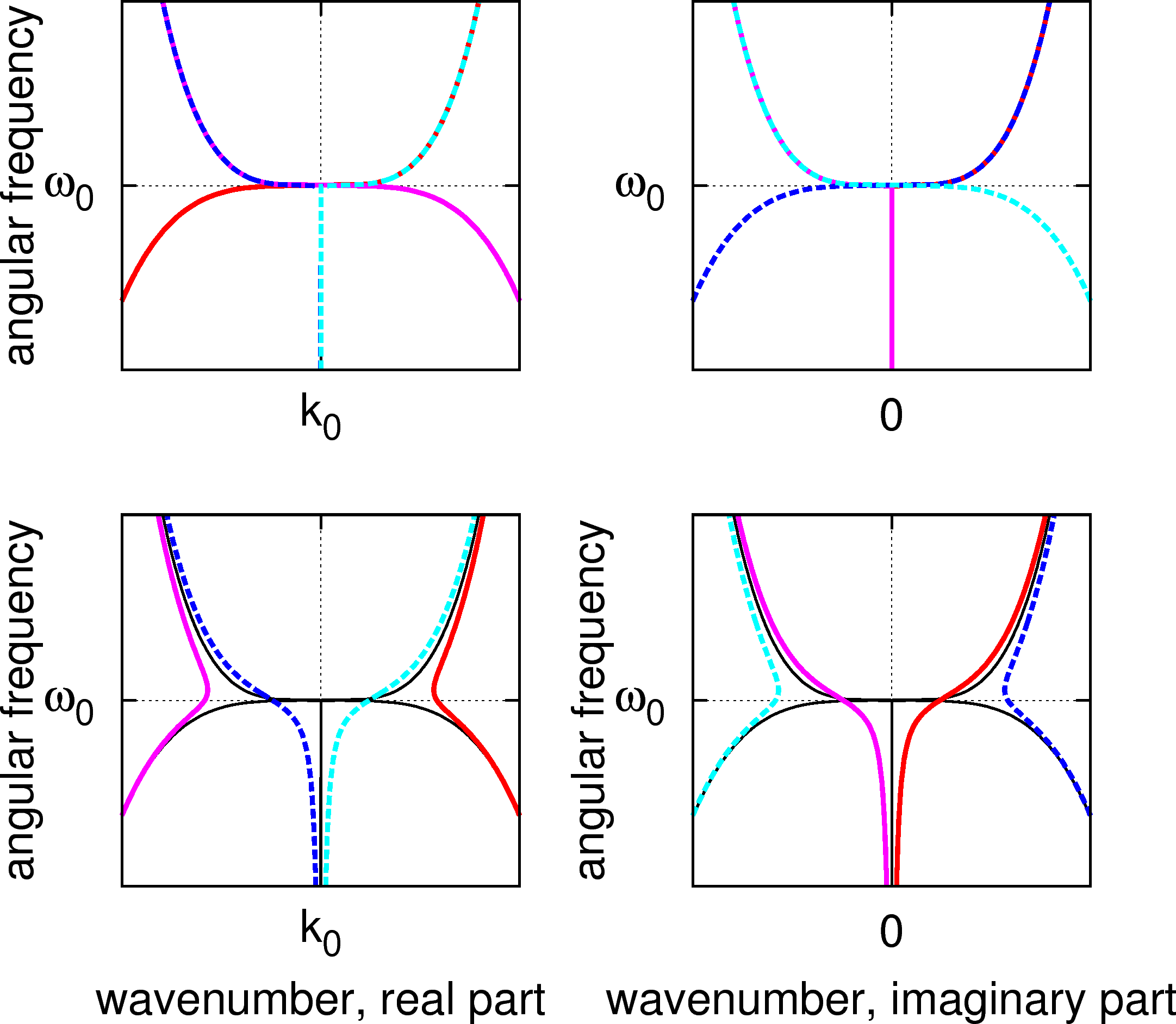}
\caption{Complex band structure $k(\omega)$ in the vicinity of a degenerate band edge (DBE) in the dispersion relation of an artificial crystal. Parameters for the plot are $\omega_0=1$, $k_0=0.5$, and $\alpha=-1$. (a) Without loss, $k(\omega)$ separates in 2 pairs of bands. Inside each pair, the bands are symmetrical with respect to $k=k_0$. (b) With loss, here with $\delta \omega / \omega_0=0.01 \imath$, $k(\omega)$ still has 2 pairs of bands that are still symmetrical with respect to $k=k_0$. Note, however, the non continuity of band sorting at the DBE when passing from the lossless case to the lossy case.}
\label{fig8}
\end{figure}

Let us now consider the case of points $(k_0,\omega_0)$ of the dispersion relation where the group velocity vanishes.
This occurs at symmetry points of the Brillouin zone, but also whenever a band is extremal at the particular point.
Figotin and Vitebskiy \cite{figotinWRCM2006} introduced the following model dispersion relation
\begin{equation}
{\cal D}(\omega, k, \mu) = \omega - \omega_0 - \alpha (k - k_0)^n = 0, n \geq 2 .
\label{eq38}
\end{equation}
If $n=2$, this is a model of a regular band edge (RBE), or more generally of a second-order stationary point, where two bands become degenerate.
If $n=3$, the model is that of a stationary inflection point (SIP) with triple degeneracy.
If $n=4$, the model is that of a degenerate band edge (BDE), implying quadruple degeneracy.
The algebraic equation (\ref{eq38}) defines at once $n$ branches of the dispersion relation, as can be seen by solving it for $k$ as a function of $\omega$ in explicit form.
Figotin and Vitebskiy considered in great detail the properties of these models in the case of lossless materials.
Let us apply to them the procedure outlined above.
The group velocity easily follows by (\ref{eq7})
\begin{equation}
v_g = n \alpha (k - k_0)^{n-1} .
\label{eq40}
\end{equation}
Since $\frac{\partial \cal{D}}{\partial k} = 0$ at point $(k_0,\omega_0)$, (\ref{eq11}) doesn't hold.
Equation (\ref{eq10}) alone, however, can be used to obtain the transformed dispersion relation as
\begin{equation}
{\cal D}(\omega, k, \mu+\delta\mu) = \omega - \omega_0 -\delta\omega - \alpha (k - k_0)^n = 0 .
\label{eq41}
\end{equation}
This equation obviously yields the band structure $\omega(k)$, but by inverting it we can in addition obtain the complex band structure $k(\omega)$ as
\begin{equation}
k = k_0 + ((\omega - \omega_0 -\delta\omega) / \alpha)^{1/n} .
\label{eq42}
\end{equation}
The last equation actually has $n$ different complex branches.
Of special interest is the departure of $k$ from $k_0$ at frequency $\omega=\omega_0$, which we still denote $\delta k$.
Its value is
\begin{equation}
\delta k = \gamma^m_n (\omega_0 F L / (2 |\alpha|))^{1/n}, m=0, \cdots, n-1
\label{eq43}
\end{equation}
with $\gamma^m_n = \exp(\mp \imath \pi (2m + 1/2) / n)$ the $n$-roots of $\mp \imath$.
The choice of sign in the previous expression must be made according to minus the sign of $\alpha$.
As a result, the spatial decay of Bloch waves close to a stationary point is proportional to the $n$-th root of the loss factor and to the $n$-th root of the local curvature of the band as measured by $1/|\alpha|$.
Figures \ref{fig6}, \ref{fig7}, and \ref{fig8} display the complex band structure for RBE, SIP, and DBE, respectively.

For a RBE (figure \ref{fig6}, plotted for $\alpha<0$), there are two complex bands associated with $\gamma^0_2$ and $\gamma^1_2$.
The two bands are symmetrical with respect to $k=k_0$.
One of them gives the dispersion of a right-propagating Bloch wave, while the other one gives the dispersion of a left-propagating Bloch wave.
Right-propagating is here defined by $\Im(k)<0$ or $\Re(k)>0$ if $\Im(k)=0$.
Left-propagating corresponds to the converse conditions.
The $k$-distance between dispersion relations with and without loss is maximum at the degeneracy frequency $\omega_0$ and equals the value in (\ref{eq41}).

For a SIP (figure \ref{fig7}, plotted for $\alpha>0$), there are 3 complex bands associated with $\gamma^0_3$, $\gamma^1_3$, and $\gamma^2_3$.
The situation is somewhat more complex as compared to the RBE case.
In the absence of loss, one band is purely real and the other two are complex conjugates.
While the real band is right-propagating, the complex conjugate bands were sorted by continuity and exchange from right- to left-propagating when going through the SIP.
In the presence of loss, a clear separation between 2 right-propagating and 1 left-propagating bands is installed.
Thus the complex band structure is not continuous when passing from $\delta \mu=0$ to $\delta \mu \neq 0$, even for an infinitesimal change.

For a DBE (figure \ref{fig8}, plotted for $\alpha>0$), there are two pairs of bands that are symmetric with respect to the vertical axis $k=k_0$.
This property is conserved when introducing loss, but similarly to the SIP case, band sorting is not continuous when passing from the lossless case to the lossy case.
Note that for both SIP and DBE, this non continuity is a consequence of the convention for sorting between left- and right-propagating Bloch waves.

The group velocity at a degenerate point subject to loss is easily found from (\ref{eq40}) and (\ref{eq43}) to be
\begin{equation}
v_g = n \alpha (\gamma^m_n)^{n-1} (\omega_0 F L / (2 |\alpha|))^{(n-1)/n} .
\end{equation}
We can obtain the limiting group velocity by considering the real part and minimizing its modulus, $v_g^L=\min_m |\Re(v_g)|$. The result is
\begin{equation}
v_g^L = n 2^{-\frac{n-1}{n}} \cos\left(\frac{(n-1)\pi}{2n}\right) |\alpha|^{\frac{1}{n}} \left( \omega_0 F L \right)^{\frac{n-1}{n}} .
\end{equation}
For $n=1$, this expression reduces to $|\alpha|$.
For $n=2$, Pedersen's expression (\ref{eq4}) is recovered.
Table \ref{tab2} summarizes the results for $n=1 \cdots 4$.

\begin{table}[!t]
\caption{Limiting group velocity for non-degenerate ($n=1$) and degenerate ($n \geq 2$) dispersion points.}
\label{tab2}
\begin{tabular}{l@{~~}l@{~~}l}
\hline
$n$ & Point type & $v_g^L$ \\
\hline
1 & linear & $|\alpha|$ \\
2 & RBE & $\sqrt{|\alpha| \omega_0 F L}$ \\
3 & SIP & $0.945 \, |\alpha|^{\frac{1}{3}} \left( \omega_0 F L \right)^{\frac{2}{3}}$ \\
4 & DBE & $1.682 \, |\alpha|^{\frac{1}{4}} \left( \omega_0 F L \right)^{\frac{3}{4}}$ \\
\hline
\end{tabular}
\end{table}

\subsection{Band gap model}

\begin{figure}[!t]
\centering
\includegraphics[width=85mm]{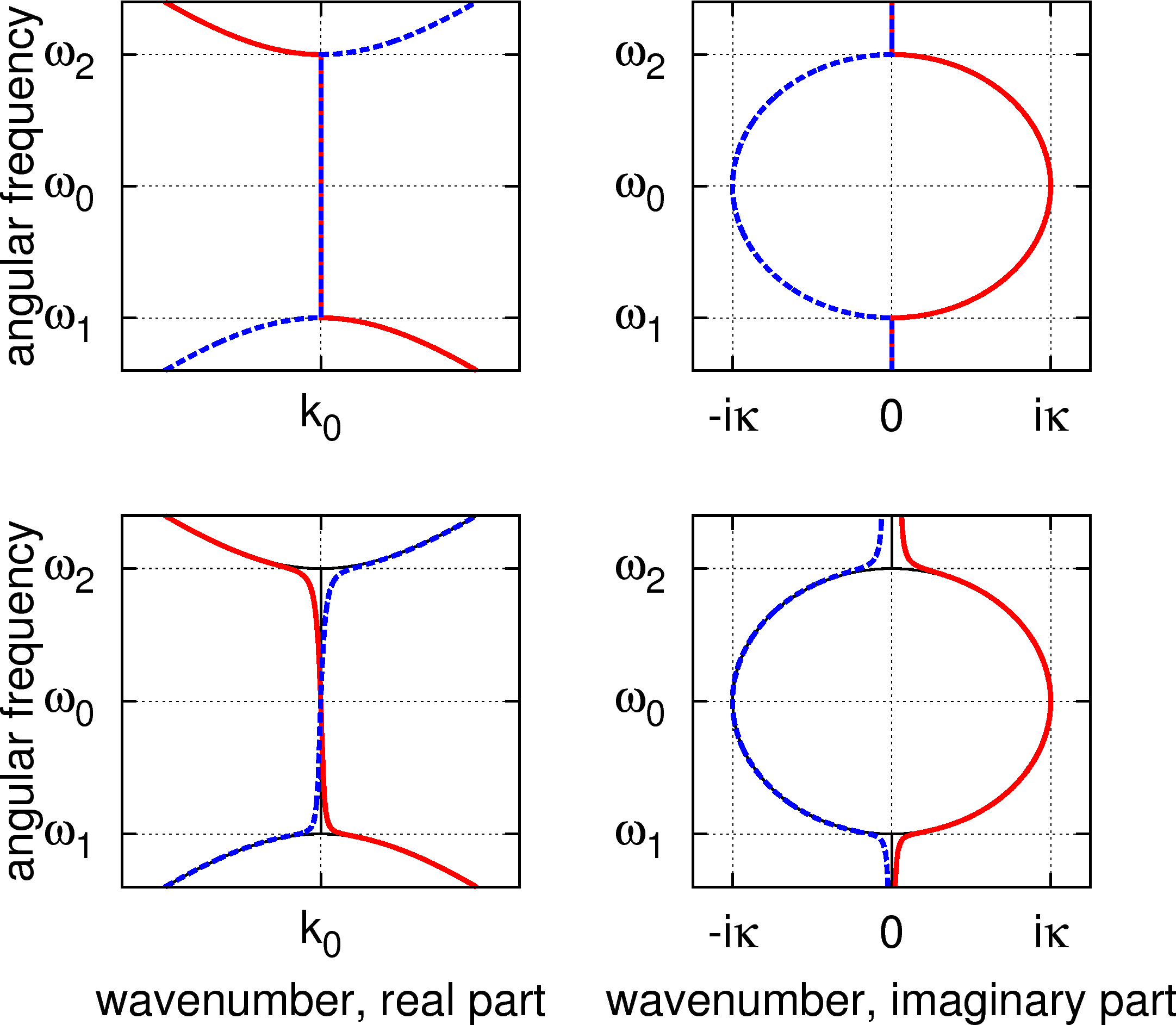}
\caption{Complex band structure for the band gap model of the dispersion relation of an artificial crystal. Parameters for the plot are $\omega_1=1$, $\omega_2=2$, $k_0=0.5$, and $\kappa=0.4$. (a) Without loss, $k(\omega)$ forms 2 bands symmetrical with respect to $k_0$. (b) With loss, here with $\delta \omega / \omega_0=0.01 \imath$, $k(\omega)$ becomes complex but the 2 bands are still symmetrical with respect to $k_0$. The right-propagating (left-propagating, respectively) Bloch waves is shown with a blue line (red line, resp.).}
\label{fig9}
\end{figure}

We propose in this section a simplified dispersion relation model that is valid around and inside a photonic or phononic band gap.
We start with an expression of the implicit dispersion relation in the absence of loss
\begin{equation}
{\cal D}(\omega, k, \mu) = \frac{(\omega - \omega_1)(\omega - \omega_2)}{(B/2)^2} - \frac{(k - k_0)^2}{\kappa^2} = 0 ,
\end{equation}
where $\omega_1$ and $\omega_2$ are the angular frequencies at the entrance and at the exit of the band gap, and $k_0$ is the band gap wavenumber (\textit{e.g.}, $k_0=\pi/a$ at the X point of the first Brillouin zone).
$\kappa$ has units of wavenumbers and measures the band gap depth.
$B = \omega_2 - \omega_1$ is the gap width.
We also introduce $\omega_0 = (\omega_1 + \omega_2) / 2$ the center of the band gap.
It is easy to check that
\begin{eqnarray}
k &=& k_0 \pm \frac{2 \kappa}{B} \sqrt{(\omega - \omega_1)(\omega - \omega_2)}, \omega \notin [\omega1, \omega_2], \\
  &=& k_0 \mp \frac{2 \imath \kappa}{B} \sqrt{|(\omega - \omega_1)(\omega - \omega_2)|}, \omega \in [\omega1, \omega_2] .
\end{eqnarray}
For angular frequency $\omega_0$, $k = k_0 \mp \imath \kappa$.
The group velocity is from (\ref{eq7})
\begin{equation}
v_g = \left( \frac{B}{2 \kappa} \right)^2 \frac{k - k_0}{\omega - \omega_0} .
\end{equation}
It can be checked by inspection that the group velocity vanishes at the entrance and at the exit of the band gap, and that it goes to infinity at the center of the band gap.

The dispersion relation can be easily complexified in the case of loss through (\ref{eq10}) and (\ref{eq25}) as exemplified with the models in the previous subsections.
We skip the details and give the main results.
At the entrance of the band gap, for $\omega = \omega_1$, we have
\begin{equation}
\delta k|_\omega \approx \pm (1+\imath) \kappa \sqrt{\omega_1 F L / B}
\end{equation}
and
\begin{equation}
v_g \approx \mp \frac{1+\imath}{2\kappa} \sqrt{\omega_1 B F L} .
\end{equation}
The limiting minimum group velocity is thus
\begin{equation}
v^L_g = \frac{\sqrt{\omega_1 B F L}}{2\kappa} .
\end{equation}
At the exit of the band gap, for $\omega = \omega_2$, we have
\begin{equation}
\delta k|_\omega \approx \pm (1-\imath) \kappa \sqrt{\omega_2 F L / B}
\end{equation}
and
\begin{equation}
v_g \approx \pm \frac{1-\imath}{2\kappa} \sqrt{\omega_2 B F L} .
\end{equation}
The limiting minimum group velocity is thus
\begin{equation}
v^L_g = \frac{\sqrt{\omega_2 B F L}}{2\kappa} .
\end{equation}

Another interesting point is the center of the band gap, for $\omega = \omega_0$.
Indeed, at this point, the group velocity is infinite in the lossless case, in connection with the classical Hartman effect.
When loss is added, we still have $k \approx k_0 \mp \imath \kappa$, but now
\begin{equation}
\Re (v_g) \approx \pm \frac{B^2}{2 \kappa \omega_0 F L} .
\end{equation}
As a consequence, loss limits the maximum group velocity at the center of the band gap, forbidding it to reach infinite values as would be expected in the lossless case.
This effect was observed experimentally and numerically in a study of tunneling in a band gap of a phononic crystal with a finite thickness.\cite{yangPRL2002}

\section{Conclusion}

We have investigated the transformation of the dispersion relation of waves propagating in artificial crystals under the addition of loss.
The results presented in this paper apply to a general anisotropic Helmholtz equation with periodic coefficients, including the cases of photonic and phononic crystals.
The consideration of an implicit form for the dispersion relation simplifies the analysis and makes it easier to construct rules for the complexification of the dispersion relation when loss is added.
As a general rule, this complexification mainly amounts to a small shift in the complex $(k,\omega)$ plane, providing first-order perturbation theory applies.
It turns out that the complex frequency shift can always be given a very simple expression that generalizes a result by Pedersen \textit{et al.}\cite{pedersenPRB2008}
Periodicity implies that at stationary points where the group velocity vanishes, first-order perturbation theory breaks down for the complex wavenumber shift.
The sole knowledge of the complex frequency shift, however, is sufficient to obtain the lossy complex band structure, as we have illustrated with the example of a family of degenerate dispersion points.
Finally, an analytical model of a Bragg-type frequency band gap was constructed and complexified for the case of loss.
For all considered models, analytical expressions were given for the limiting group velocity, i.e., for the lower bound of the group velocity that is imposed by loss.
These analytical expressions are in excellent agreement with the transformations of the complex band structure of phononic crystals subject to material loss that were observed numerically.\cite{moiseyenkoPRB2011}
Furthermore, and although we presented our approach in the frame of periodic media (photonic and phononic crystals), it could be useful to analyze the influence of loss on the dispersion of waveguides and more generally of structures presenting dispersion resulting from the spatial distribution of materials.
For instance, plasmonic waveguides in the form of silver nanorods were shown to present mode transformations similar to the RBE model.\cite{davoyanPN2011}

\begin{acknowledgments}
Financial support from the European Community's Seventh Framework program (FP7/2007-2013) under grant agreement number 233883 (TAILPHOX) is gratefully acknowledged.
VL acknowledges the support of the Labex ACTION program (contract ANR-11-LABX-01-01).
\end{acknowledgments}

\bibliographystyle{apsrev}
\bibliography{../../vince}

\end{document}